\documentstyle[aps,twocolumn,epsf]{revtex}
\begin{document}
\title{Hydrodynamic Excitations of Trapped Fermi Gases}
\author{Georg M.\ Bruun and Charles W.\ Clark}
\address{Electron and Optical Physics Division, 
National Institute of Standards and Technology,
Technology Administration, 
US Department of Commerce, Gaithersburg, 
Maryland 20899-8410}
\maketitle
\begin{abstract}
We discuss collective excitations of a trapped dilute Fermi gas within a 
hydrodynamic approximation. Analytical results are derived for both high-
and low-temperature limits  and are applied to $^{40}$K and $^6$Li systems
of current experimental interest. We identify spectral signatures which can be
used to detect the onset of Fermi degeneracy. Also, we find an interesting class of
internal excitations with an unusual spectrum. Some of our results are relevant 
to the case of trapped bosons as well. Our analysis suggests several experiments
 which address fundamental problems of collective motion in quantum fluids. 
\end{abstract}
A new era in low-temperature physics was heralded by the production in 1995 
of Bose-Einstein condensation (BEC) 
in the alkali gases $^{87}$Rb, $^{23}$Na, and $^7$Li.~\cite{BEC} 
These dilute-gas systems are governed by atomic interactions
that are very well characterized, so they provide systems
in which quantum many-body physics
can be explored from first principles. They have also been used for a remarkable
variety of experiments on coherent matter wave generation and propagation, and have stimulated
much experimental and theoretical work.\cite{Edwards1}   

Recently, impressive experimental results concerning the quantum degenerate regime 
of a dilute gas of trapped fermionic $^{40}$K atoms have been presented~\cite{DeMarcoScience}
thereby enabling fundamental investigations of quantum-degenerate Fermi systems with
controlled particle interactions. Furthermore, reports of magneto-optical trapping of 
$^6$Li have been presented recently.\cite{Mewescon} 
Several theoretical results concerning the equilibrium properties of 
such a gas have been presented.~\cite{Butts,Schneider,BruunN,StoofBCS,BruunBCS}

In the present paper, we  examine the spectroscopy of collective modes of excitation
of the Fermi gas. The study of this spectrum has been fruitful in studies of BEC.
 These modes correspond to sound waves in a homogeneous gas, but for a confined gas
their spectrum is discrete. They can be observed as shape oscillations of the 
atomic cloud, induced by perturbations of the trapping potential.  Such measurements have 
been carried out for Bose gases, for cases of a nearly pure condensate \cite{Jin,Mewes}, for a 
partially-condensed gas at temperatures comparable to the BEC transition
temperature $T_0$, \cite{JILAfinite,StamperKurn} and in the normal state \cite{StamperKurn}. 
For $T < 0.6 T_0$ the observed collective excitation frequencies are in good agreement with 
predictions of first--principles mean field theory
\cite{Edwards2,Dodd1}, and, for gases with more than a few thousand atoms, they agree
well with the predictions of a simpler hydrodynamic theory,~\cite{Stringari,CastinDum} which
resembles the approach we develop in this paper. 
 
As is well known \cite{Pines}, the collective modes of a quantum fluid
admit simple descriptions in the collisionless and the hydrodynamic regimes.
When the lifetime $\tau$ of the quasiparticles is much longer
than the characteristic period of motion ({\it{i.e.}} $\omega\tau\gg 1$
for atoms in a trap of frequency $\omega$), 
there are few scattering events per sound oscillation, and the restoring
force is due the self-consistent mean field of the gas. 
Wave motion encountered in this limit is designated ''zero sound''. 
For the hydrodynamic regime $\omega\tau\ll 1$, on the other hand, collisions
ensure local thermodynamic equilibrium.  To attain the
hydrodynamic regime in an ultracold gas of fermionic atoms, it
is necessary to trap at least two different hyperfine states,
since the ($p$-wave) interaction between atoms in the same hyperfine state
is completely suppressed below $T\simeq 100\mu$K.~\cite{DeMarco} 
Experimentally, trapping two hyperfine states has been found to provide
the mechanism for collisional rethermalization needed
for evaporative cooling.~\cite{DeMarcoScience}  This paper treats 
the case of a two-component Fermi gas in the hydrodynamic regime.
We present a systematic analysis of the modes in the Fermi degenerate 
regime $T\ll T_{\rm F}$ and in the classical limit $T\ge{\mathcal{O}}(T_{\rm F})$. 
Our analytical results for the density dependence of collective excitation frequencies
 suggest a means of detecting the onset of Fermi degeneracy in the 
gas at low $T$. We relate our results to current experiments on $^{40}$K and $^6$Li.

The problem of collective modes of spatially confined two-component Fermi systems is 
well-known in the field of  nuclear physics. However, due to the relatively 
few nucleons in a typical nucleus, these systems tend to be most appropriately described in the 
zero sound limit contrary to the situation described by the present paper.~\cite{Goeke}

In the limit of low energy/long wavelength excitations, one can treat the dynamics of a quantum
gas by a semiclassical approximation, via the Boltzmann equation 
for the function $f({\mathbf{r}},{\mathbf{p}},t)$, which describes the distribution of 
particles of mass
 $m$ with position ${\mathbf{r}}$ and momentum ${\mathbf{p}}$.
In the hydrodynamic regime, collisions ensure local equilibrium with a given mean velocity 
${\mathbf{u}}({\mathbf{r}},t)$, temperature $T({\mathbf{r}},t)$, and mass density 
 $\rho({\mathbf{r}},t)$, since they conserve net momentum, energy and particle number. 
Assuming that the gas is otherwise ideal, we look for solutions to the Boltzmann equation 
that yield the Fermi-Dirac distribution
 $f({\mathbf{r}},{\mathbf{p}},t)=\{\exp[\beta({\mathbf{r}},t)\xi({\mathbf{r}},t)]+1\}^{-1}$
with $\xi({\mathbf{r}},t)=[{\mathbf{p}}-m{\mathbf{u}}({\mathbf{r}},t)]^2/2m-
\mu_{\rm F} ({\mathbf{r}},t)$ and $\beta({\mathbf{r}},t)=1/k_{\rm B} T({\mathbf{r}},t)$.
In this limit, the linearized momentum and energy conservation laws are:
\begin{eqnarray}\label{Pconservation}
\rho_0 ({\mathbf{r}})\partial_t {\mathbf{u}}({\mathbf{r}},t)=
-\nabla P({\mathbf{r}},t)+
\rho({\mathbf{r}},{\mathbf{p}},t) {\mathbf{f}} \\ \label{Econservation}
\partial_t P({\mathbf{r}},t)=-\frac{5}{3}\nabla[P_0({\mathbf{r}})
{\mathbf{u}}({\mathbf{r}},t)]+
\frac{2}{3}\rho_0 ({\mathbf{r}})
{\mathbf{u}}({\mathbf{r}},t){\mathbf{f}}.
\end{eqnarray}
with ${\mathbf{f}}=-\nabla V_{pot}({\mathbf{r}})/m$. Here, $V_{pot}({\mathbf{r}})$
 is the trapping potential and  
 $\rho({\mathbf{r}},t)=\rho_0 ({\mathbf{r}})+\delta\rho({\mathbf{r}},t)$, 
$P({\mathbf{r}},t)=P_0 ({\mathbf{r}})+\delta P({\mathbf{r}},t)$ with  $\rho_0 ({\mathbf{r}})$
 and $P_0 ({\mathbf{r}})$ being the equilibrium mass density and pressure respectively. 
As we are interested in modes for which the densities of the two states oscillate 
in phase, we have defined the total density  
$\rho ({\mathbf{r}},t)\equiv\sum_\sigma\int d^3k/(2\pi)^3f({\mathbf{r}},{\mathbf{p}},t)=
\Sigma_\sigma\rho_\sigma({\mathbf{r}})$
 where $\sum_\sigma$ denotes the sum over the two trapped hyperfine states.
Likewise, $P({\mathbf{r}},t)$ and ${\mathbf{u}}({\mathbf{r}},t)$ denote the total 
pressure and mean velocity associated with both species. We take the number of 
atoms in each hyperfine state to be the same, so that there is a single chemical potential and 
 $\rho({\mathbf{r}})=2\rho_\sigma({\mathbf{r}})$. The "spin"-modes, 
for which the two components of the gas do not move in phase, will be purely 
diffusive in the hydrodynamic regime, as the collisions suppress any difference 
in the local velocity of the two components.~\cite{Vichi} Using the continuity equation,
$\partial_t\rho({\mathbf{r}},t)=-\nabla[\rho({\mathbf{r}},t)
{\mathbf{u}}({\mathbf{r}},t)]$, Eqs.
(\ref{Pconservation})-(\ref{Econservation})
can be written as a closed equation for the velocity field \cite{Griffin}:
\begin{equation}\label{uequation}
\partial^2_t{\mathbf{u}}=\frac{5P_0}{3\rho_0}\nabla(\nabla{\mathbf{u}})
+\nabla({\mathbf{uf}})+\frac{2}{3}{\mathbf{f}}(\nabla{\mathbf{u}}).
\end{equation}
Hereafter, we take the trap to be an isotropic harmonic oscillator, i.e.\ 
$V_{pot}=m\omega^2r^2/2$.
The hydrodynamic equations are then solved in several limiting cases. We first
consider the limit $T\ll T_{\rm F}$, where the pressure is a function only
of the density and we invoke the semiclassical (Thomas-Fermi)
approximation to the equilibrium density distribution,~\cite{Butts}
\begin{equation}
\rho_0(r) = \frac{8Nm}{\pi^2 r_{\rm F}^6}
\left(r_{\rm F}^2-r^2\right)^{3/2},
\label{scdensity}
\end{equation}
where $N$ is the total number of atoms in the trap, and $r_{\rm F} =
\sqrt{2\mu_{\rm F}/m\omega^2}$, with $\mu_{\rm F} = (3N)^{1/3}\hbar\omega$.
From Eq.(\ref{Econservation}) we obtain
$P_0/\rho_0=\omega^2(r_{\rm F}^2-r^2)/5$, so Eq.(\ref{uequation}) becomes
\begin{equation}
\partial_t^2{\mathbf{u}}=
\frac{\omega^2}{3}\nabla\left[(r_{\rm F}^2-r^2)\nabla{\mathbf{u}}\right]-
\omega^2\nabla({\mathbf{ur}}).
\end{equation}
As the right hand side of this equation is a pure gradient, the sound modes can be 
found by considering potential flow. Equation (\ref{Pconservation}) and the
continuity equation then decouple, to yield
\begin{equation}\label{zeroTeqn}
\partial^2_t\delta\rho=\omega^2\left[
1-\frac{r}{3}\partial_r + \frac{1}{3}(r_{\rm F}^2 -
r^2)\nabla^2\right]\delta\rho.
\end{equation}
Eq.(\ref{zeroTeqn}) clearly separates in spherical polar coordinates
($r,\theta,\phi$), so we invoke the eigenfunction expansion,
\begin{equation}
\label{separation}
\delta\rho({\mathbf{r}}, t)=
\sum_{nlm}e^{i\omega_{nl}t}
 \delta\rho_{nl}(r) Y_{lm}(\theta,\phi) ,
\end{equation}
where $n$ is an eigenvalue index, $\omega_{nl}$ a sound eigenfrequency, and
$\delta\rho_{nl}(r)$ the corresponding density fluctuation eigenfunction (with a normalization to 
be determined by boundary conditions). Substituting Eq.(\ref{separation}) into Eq.(\ref{zeroTeqn}) gives
an ordinary differential equation with three regular singular points, and the condition 
that $\delta\rho_{nl}(r)$ be finite at all $r$ yields the eigenfrequencies~\cite{Amoruso}
\begin{equation}
\omega_s=\frac{2\omega}{\sqrt{3}}\sqrt{n^2+2n+ln+3l/4}
\label{spectrum}
\end{equation}
and the (unnormalized) eigenfunctions
\begin{equation}
\delta\rho_{nl}(r) = 
r^l \sqrt{r_{\rm F}^2 - r^2}\   _2F_1[-n, l+2+n, l+3/2; (r/r_F)^2]
\label{normalmodes}
\end{equation}
where $n = 0, 1, 2, . . .,$ and $_2F_1$ is the usual hypergeometric
function \cite{Abramowitz}, which is equivalent to the Jacobi polynomial
 $P_n^{(l+1/2,1/2)}$. For $l=0$, the solutions reduce to
\begin{eqnarray}
\delta\rho(r)_{n0}\propto\sin[(2n+2)\arccos(r/r_F)]/r\nonumber\\
\propto\sqrt{1-r^2/r_F^2}U_{2n+1}(r/r_F)/r
\end{eqnarray}
with $U_n(x)$ being the Chebyshev polynomials. Note that the zero-frequency 
solution, $\omega_{00}=0$, is associated with a mode,  
$\delta\rho_{00}(r) = (1-r^2/r_F^2)^{1/2}$, which corresponds to an
infinitesimal change in the number of particles in
the system [see Eq.(\ref{scdensity})]. It is therefore not a physical sound
mode.  All other modes described by Eqs.(\ref{spectrum}-\ref{normalmodes})
are particle conserving [i.e.\ $\int d^3r\delta\rho_n(r)=0$], as required. The
lowest frequency for a given $l$ is  $\omega_{l0}=\sqrt{l}\omega$, a result first
derived by Griffin \textit{et al}.~\cite{Griffin} The $\omega_{10}=\omega$ mode 
corresponds to the lowest center-of-mass oscillation.  As noted by Griffin  
\textit{et al.}\cite{Griffin}, only modes for which 
 $\nabla {\mathbf{u}}\neq constant$ depend on the statistics (i.e.\ $P_0/\rho_0$) of 
the atoms as can be seen directly 
from Eq.(\ref{uequation}).  By calculating  the complete sound spectrum, we
have found these modes. From $\nabla^2[r^lY_{lm}(\theta,\phi)]=0$, we see that 
for any angular momentum, the lowest observed mode does not depend on the statistics 
of the gas whereas the higher ones ($n>1$ for $l=0$ and $n>0$ for $l>0$) do.

We now consider the high $T$  regime where  the de Broglie wavelength 
is much smaller than the interparticle spacing and the gas approaches the
classical limit. However, we assume sufficiently low temperatures such that the
density is high enough for the hydrodynamic approximation to be valid.
To find the sound modes in this regime, we need to return to Eq.(\ref{uequation})
as the pressure is now not related to the density in a simple way. For
$l=0$, the velocity field is irrotational and we write 
 ${\mathbf{u}}=\nabla [p_n(x)]$, where $p_n$ is a polynomial in $x=r/r_{\rm TF}$ with 
 $r_{\rm TF}=\sqrt{k_{\rm B}T/m\omega^2}$. Using $P_0/\rho_0=k_{\rm B}T/m$, we 
find the sound frequencies
\begin{equation} \label{classical}
\omega_{0n}=\omega\sqrt{10n/3+4}
\end{equation}
with $n=0,1,2,\ldots$. The corresponding density fluctuations are given by
\begin{equation} 
 \delta\rho(r)\propto (2/r+\partial_r)[e^{-\beta
m\omega^2r^2/2}\partial_rp_n(r/r_F)]
\end{equation}
where $\partial_xp_n(x)=x[b_0+b_2x^2+\ldots+b_{2n}x^{2n}]$.  As expected, we
recover the $\omega_s=2\omega$ mode which, since $\nabla {\mathbf{u}}=constant$, 
is independent of the particular form of $P_0/\rho_0$. However, the higher modes 
($\nabla {\mathbf{u}}\neq constant$) do depend on the statistics 
as can be seen by comparing Eq.(\ref{spectrum}) with $l=0$ and
Eq.(\ref{classical}). For general angular 
momentum $l$, the analysis is somewhat complicated by the fact that the
velocity modes are not irrotational. Writing 
 ${\mathbf{u}}=-\nabla\psi+\nabla\times{\mathbf{A}}$ and 
$2{\mathbf{r}}\nabla{\mathbf{u}}/3=-\nabla\tilde{\psi}+\nabla\times\tilde{{\mathbf{A}}}$,
 we obtain from Eq.(\ref{uequation}) the coupled equations:
\begin{eqnarray}
\partial_t^2\psi&=&\frac{5k_{\rm
B}T}{3m}\nabla^2\psi-\omega^2(\nabla\psi-\nabla\times{\mathbf{A}})
{\mathbf{r}}-\omega^2\tilde{\psi}\\
\partial_t^2(\nabla\times{\mathbf{A}})&=&-\omega^2\nabla\times\tilde{{\mathbf{A}}}.
\end{eqnarray}
They are solved  by writing $\psi_{nl}=p_{nl}(x)x^lY_{lm}(\theta,\phi)$ with
 $p_{nl}=a_{0l}+a_{2l}x^2+\ldots+a_{2nl}x^{2n}$. The 
spectrum is
\begin{eqnarray}\label{classicall}
\frac{\omega_s^2}{\omega^2}&=&\frac{1}{2}\left[\frac{5}{3}(l+2n+2/5)\right.\nonumber\\
&\pm&\left. \sqrt{(l+2n+2/5)^225/9-l(l+1)8/3}\phantom{\frac{5}{3}}\right]
\end{eqnarray}
with $n=0,1,2,\dots$. The result $\omega_s=\sqrt{l}\omega$  for the lowest
mode for a given $l$ independent of statistics is recovered as expected.
Again, by comparing Eq.(\ref{spectrum})  and Eq.(\ref{classicall}) we see
that the higher modes depend on the functional form of the density profile. 
 The corresponding density fluctuations obtained from the continuity
equation are 
 $\delta\rho_{nl}\propto\exp(-x^2/2)
x^l[\tilde{a}_{0l}+\tilde{a}_{2l}x^2+\ldots+\tilde{a}_{2nl}x^{2n}]Y_{lm}(\theta,\phi)$.
 Using this, it is easy to show that the density fluctuations are particle
conserving. The high $T$ sound frequencies are independent of $T$ which is a special
property of the harmonic trap. One should note that the hydrodynamic approximation 
starts to break down for the high lying modes ($n$ large) when the characteristic wavelength
of these  modes becomes comparable to the mean free path.

The eigenfrequencies given by subtracting the square root in Eq.(\ref{classicall})
 (for $n\ge 1$) represent an interesting class of solutions. These modes have
\emph{decreasing} frequency with increasing wave number which is characteristic for 
the so-called internal waves. Internal waves are present in general when the equilibrium 
density profile is stratified by an external force. They are, contrary to ordinary sound, 
essentially driven by the external force giving rise to the unusual relation 
between frequency and wave number.~\cite{Lighthill} A detection  of these 
modes would be a very useful experimental verification of a general result in the 
theory of hydrodynamics.

We will now turn to possible experimental implications  of the results
presented. As mentioned above, in order to achieve the hydrodynamic limit
 $\omega\tau\ll 1$ one should trap the relevant fermionic atoms in two
hyperfine states. We need to examine the effect of the Fermi blocking on the 
scattering rate
of the atoms. In the presence of a Fermi sea, the low energy lifetime  $\tau$ of a
quasiparticle for low temperatures is $\tau_{class}/\tau\propto (T/T_{\rm F})^2$
where $\tau_{class}^{-1}\simeq \rho_\sigma\sigma_0\langle v\rangle/m$ 
is the classical scattering rate, $T_F=\mu_F/k_B$, and $\langle v\rangle$  is the thermal
velocity of the atoms. For the energies relevant for the present problem,
only $s-$wave scattering between atoms in different hyperfine states is important
and $\sigma_0=4\pi a^2$ where  $a$ is  the $s-$wave scattering length. 
 The $(T/T_{\rm F})^2$ factor reflects the increasing Fermi blocking of the
scattering states with decreasing temperature.~\cite{Pines} 
For $T\ge\mathcal{O}(T_F)$, the Fermi blocking
becomes negligible, and $\tau^{-1}\simeq\tau^{-1}_{class}$. As $\langle v\rangle\propto T^{1/2}$
and $\rho(0)\propto T^{-3/2}$, we have $\tau^{-1}_{class}\propto T^{-1}$.
Thus, both for $T\rightarrow 0$ and $T\rightarrow \infty$ the gas is in the
collisionless regime and we need to examine in which region of the phase diagram the
hydrodynamic description presented in this paper is valid. We will concentrate  
on the two atoms  $^6$Li and $^{40}$K and assume a trapping frequency of 
 $\nu=\omega/2\pi=144$Hz approximating 
typical experimental conditions. Of course, by altering the trapping
frequency one can change the regions  where the hydrodynamic approximation is valid.

For $^6$Li, the scattering length is $a\simeq -2160a_0$ where $a_0$ is the
Bohr radius.~\cite{Abraham} This large value should mean that the
hydrodynamic description is valid in large regions of the phase diagram. 
Indeed, using $\nu=144$Hz and assuming that there is $10^6$ particles 
trapped in each hyperfine state, we obtain that the hydrodynamic
approximation is accurate down to temperatures $T\sim 0.1T_{\rm F}$. For such 
low temperatures, the gas is in the Fermi degenerate regime and we expect 
Eqs.(\ref{spectrum}-\ref{normalmodes}) to give a good description of the
sound modes. As $T$ is increased to $T\sim  T_F$, $P_0/\rho_0$ approaches the 
classical form and the sound spectrum is given by Eq.(\ref{classicall}).
Since the scattering length for $^6$Li is so large, the collisionless limit is only
reached for a relatively low density ( $T\gg T_F$ for N=$2\times 10^6$). Thus, for 
a wide temperature range the hydrodynamic approximation is valid and  the 
sound spectrum should be independent of $T$ and given by
Eq.(\ref{classicall}). The onset of Fermi degeneracy for $T<T_F$ should show up in 
a $T$ dependence of the sound spectrum as it changes from Eq.(\ref{classicall}) to
Eq.(\ref{spectrum}) with decreasing $T$ reflecting the change in the functional form of 
 $P_0/\rho_0$. Eventually, for  $T\gg T_F$ we have $\omega\tau\sim 1$ and the damping 
of the modes becomes large \cite{Pines,Guery,Kavoulakis}. For higher $T$, the gas is in the 
collisionless regime. In this regime, the frequencies approach  the single particle
spectrum, e.g.\  $\omega_s\simeq 2n\omega$ for $l=0$, as the effect of the interactions 
decrease with decreasing density~\cite{Vichi}. 
Of course, by trapping more/fewer atoms one can increase/decrease 
the temperature region where the hydrodynamic description is valid. 

The spectrum given by Eq.(\ref{spectrum}) coincide  with the $T=0$ 
collective mode spectrum in the BCS state.~\cite{Baranov} Thus, the density-fluctuation 
spectrum in the hydrodynamic limit of the normal phase and in the superfluid phase is the same. 
This generalizes  the well-known result for homogeneous systems,~\cite{Schrieffer}
to the case of trapped systems. The effect of the superfluid correlations on the various
 collective modes in general has been examined in detail for homogeneous systems.~\cite{Vaks}

For $^{40}$K, the situation is very different since the scattering length 
 $152a_0\leq a\leq542a_0$ is much smaller.~\cite{Bohn} Therefore, in order to reach the
 hydrodynamic regime one needs to trap at least $\sim 10^8$ atoms in each hyperfine state.
Alternatively, one could increase the scattering rate by adjusting an external magnetic
field to achieve a Feshbach resonance. This latter procedure is also expected to be necessary
to obtain a large and negative scattering length in order to get  an
experimentally realistic transition temperature $T_c$ for the BCS transition.
Therefore, although the treatment of the sound modes put forward in this paper does not describe
the present experimental situation  for  $^{40}$K, one might expect it to be relevant
for future experiments.

Of course, Eq.(\ref{classical}) and Eq.(\ref{classicall}) are still valid
for high temperatures for a bosonic gas in the hydrodynamic regime. As an additional
result, we therefore suggest  that one  should also be able to observe the transition from the
classical to the quantum regime for a Bose gas by looking at the change in the sound
spectrum. The modes for high $T$ are presented in this paper whereas  the modes in the BEC 
limit are described in ref.\ \cite{Stringari}.

For an anisotropic gas, a calculation of the higher ($\nabla{\mathbf{u}}\neq constant$)
 modes is somewhat cumbersome algebraically. However, although 
the spectrum will be different, the qualitative behavior is the same as for the
isotropic case presented here: the modes are still independent of $T$ in the
classical limit and one should be able to detect the onset of Fermi degeneracy by
looking for a $T$-dependence of the modes. Thus, the main conclusions of the paper
remain valid for the anisotropic case.

In conclusion, we have analyzed the collective modes of a dilute trapped
Fermi gas in the hydrodynamic limit. Analytical results both in the low and high  $T$ limit for  modes 
which depend on the density profile of the cloud were presented. This lead to the proposal of a 
straightforward method to detect the onset of
 Fermi degeneracy. We found an interesting class of internal modes which have decreasing 
frequency for increasing wave number.   
 Furthermore, the analysis presented for the high $T$ spectrum should
also be valid for the case of trapped bosons. Our analysis should have direct relevance to
current experiments on $^6$Li atoms and to possible future experiments on 
 $^{40}$K. 

We appreciate useful discussions with D.\ J.\ Allwright, H.\ Ockendon, and D.\ S.\ Jin.

\end{document}